\documentstyle[eqsecnum,aps,prd]{revtex}
\newcommand{\be}{\begin{equation}}
\newcommand{\ee}{\end{equation}}
\newcommand{\bea}{\begin{eqnarray}}
\newcommand{\eea}{\end{eqnarray}}
\newcommand{\eqn}[1]{(\ref{#1})}
\newcommand{\wsig}{\widehat{\Sigma}}
\newcommand{\wsigp}{\widehat{\Sigma}'}
\newcommand{\wa}{\widehat{A}}
\newcommand{\wap}{\widehat{A}'}
\newcommand{\wb}{\widehat{B}}
\newcommand{\wbp}{\widehat{B}'}
\newcommand{\wc}{\widehat{C}}
\newcommand{\wcp}{\widehat{C}'}
\newcommand{\slp}{/\!\!\!p}

\begin{document}
\setlength{\unitlength}{1mm}

{\hfill hep-ph/9805428}

{\hfill DSF 14/98}\vspace*{2cm}

\begin{center}
{\Large \bf Wave Function Renormalization at Finite Temperature}
\end{center}

\bigskip\bigskip

\begin{center}
{\bf S. Esposito}, {\bf G. Mangano}, {\bf G. Miele} and {\bf O. Pisanti}
\end{center}

\vspace{.5cm}

\noindent
\begin{center}
{\it Dipartimento di Scienze Fisiche, Universit\'{a} di Napoli "Federico II",\\
and \\
INFN, Sezione di Napoli, Mostra D'Oltremare Pad. 20, I-80125 Napoli,
Italy}\\
\end{center}
\bigskip\bigskip\bigskip

\begin{abstract}
We present a derivation of the medium dependent wave function
renormalization for a spinor field in presence of a thermal bath. We show
that, as already pointed out in literature, projector operators are not
multiplicatively renormalized and the effect involves a non trivial spinor
dependence, which disappears in the zero temperature covariant limit. The
results, which differ from what already found in literature, are then
applied to the decay of a massive scalar boson into two fermions and to the
$\beta$--decay and crossed related processes relevant for primordial
nucleosynthesis.
\end{abstract}
\vspace*{2cm}

\begin{center}
{\it PACS number(s):98.80.Cq, 11.10.Wx, 95.30.Cq}
\end{center}

\baselineskip=.8cm

\section{Introduction}
The large amount of new and precise data on the structure of the universe,
which will be provided by the new generation of experiments \cite{Schramm},
represents a new exciting source of information for particle physics too.
In particular, new measurements on the primordial abundance of light
elements will probably provide a severe arena where to test new models for
fundamental interactions. In this respect, the theoretical predictions on
Big Bang Nucleosynthesis (BBN), which are considered one of the great
successes of the hot Big Bang theory, must be refined in order to reach the
same level of precision of the new experimental data.

One of the necessary improvement in the BBN algorithm is to take into
account that all the nucleosynthesis reactions occur in a plasma of
photons, electron--positron pairs and neutrinos whose temperature varies,
in the relevant epoch for BBN, in the range $0.1 \div 10$ MeV (nucleons and
nuclei due to their large mass with respect to temperature may be safely
considered at $T=0$). This observation led several authors
\cite{Dicus,Cambier} to add finite temperature contributions to weak
transitions like $n + \nu_e \leftrightarrow p + e^-$ and crossed phenomena,
due to thermal radiative electromagnetic corrections. The size of these
effects on relevant cosmological observables, like the ratio $n/p$, which
is the key parameter for the evaluation of the abundances of light nuclei,
as $D$, $~^3 \mbox{He}$ or $~^{4}\mbox{He}$, has been estimated to be not
larger than few percent \cite{Dicus,Cambier}. However, since the new
experimental data will presumably be sensible to corrections of this order
of magnitude, the study of the thermal effects is certainly physically
well-motivated.

Although there is a general agreement in the literature on how to compute
finite temperature effects on phase-space, vertex, mass corrections and
photon emission/absorbtion \cite{Dicus}-\cite{Kobes}, the finite
temperature wave function renormalization still remains an open problem.
Since Refs \cite{Dicus,Cambier}, where unfortunately the problem was not
addressed in the proper way, several approaches have been proposed in
literature \cite{Donoghue}-\cite{Sawyer}. They all agree on the idea of
using finite temperature Dirac spinors to obtain the corresponding
effective projection operator. The final results, however, striking differ
\cite{Chapman}.

In this paper we approach this problem in a different but straightforward
way. Our results, obtained in section II, differ from what has been
obtained in \cite{Donoghue} for an additional term which is due to a
peculiar property of the heat bath. It has a different spinorial structure
and spoils the multiplicative character of renormalization of
wave--function. The presence of such a term was first recognized in
\cite{Sawyer}, but we disagree on its explicit form. We then perform a
comparative analysis of the several approaches and, in section III, discuss
some physical implications, like the change on the decay process of a heavy
scalar boson into lepton--antilepton pairs, $H
\rightarrow l^+ l^-$,
and on $\beta$--decay $n \rightarrow p+e^- +
\overline{\nu}_e$
and crossed related processes. Finally, in section IV we
give our conclusions.

\section{Finite temperature projection operators}

The ambiguity in the wave function renormalization at finite temperature is
basically due to the lack of Lorentz covariance. The thermal bath
introduces a preferred frame, as the one in which the time-like four vector
generalizing the non--relativistic temperature parameter takes the form
$(\beta,\vec{0})$. In this rest frame, the equilibrium particle
distribution are the usual Bose-Einstein or Fermi-Dirac expressions, with
$\beta$ the inverse temperature. The very definition of a particle state,
which corresponds in zero temperature quantum field theory to an
irreducible representation of the Poincar\`{e} group, becomes tricky and
different points of view may naturally lead to quite different results for
the medium dependent wave-function renormalization, basically due to
different choices for the renormalized fields. As noticed in \cite{Sawyer},
however, the non-local, momentum dependent effects of the medium imply that
such a field, as a local field, could even be not defined.

We will take, as in \cite{Sawyer}, quite a different point of view,
identifying the particle states as corresponding to the energy poles in the
propagator. In the covariant case, for a spinor field, the propagator will
always develop poles in the Lorentz invariant $\slp$, or equivalently in
$p^2$, the effect of the interactions being the shift of the pole and a
change in the corresponding residue. For a particle propagating in a
thermal bath, the interactions with the surrounding plasma spoil, in
general, this property, and the location of the pole cannot be represented
by an invariant statement. The mass shift would in fact acquire a momentum
(and frame) dependence, which is to say that particles propagating with
different speeds in the bath rest frame will acquire, in general, different
inertia due to the interactions with the plasma (spatial dispersion).

Notice that, unlike the zero temperature case, the effect of the wave
function renormalization, being momentum dependent as well, cannot be
absorbed via the introduction of a local countertem in the bare lagrangian
density. It therefore represents a genuine physical effect of the medium on
the propagating particles.

It is not the aim of this paper to investigate the fundamental aspects
sketched above, but rather to discuss a prescription for evaluating the
wave-function renormalization effects. They cannot, nevertheless, be hidden
under the carpet, and they render our point of view, at least with our
present understanding, only a reasonable way of dealing with this problem.

As already mentioned, our starting point will be to identify the particle
states as corresponding to the energy poles in the field propagator
$G(E,\vec{p})$, which can be perturbatively evaluated taking into account
the effects of interactions with the surrounding medium. Actually, only the
poles which are perturbatively close to the free particle ones will be
considered, and not, for example, the hole branches described in
\cite{Weldon}. The particle state will be therefore characterized by a
momentum $\vec{p}$ and a new dispersion relation $E=E(\vec{p})$ given by
the new position of the pole. The wave function renormalization, which is
in general momentum dependent, can then be read off by evaluating the
residue of $G$ at the pole. This prescription has been already considered
by Sawyer \cite{Sawyer}, but obtaining quite different results from ours.
We will comment on this later. The relevant quantity which appears in the
expressions of scattering cross sections and decay rates are the projection
operators on positive ($+$), or negative ($-$), energy states $\Lambda^{\pm}$.
The fact that they modify with respect to the free field case is of course
due to the wave function renormalization, and is the way this effect shows
up in interaction processes. In what follows, we will report our results
directly in terms of these quantities. Hereafter all calculations are
performed in the medium rest frame.

In order to develop a general approach to the finite temperature wave
function renormalization, let us first consider the simple and well-known
zero temperature case, where the problem can be easily solved using
covariance, and looking for the pole shift in $\slp$. If $\Sigma = \Sigma_1
\slp + \Sigma_0$
is the self-energy produced by radiative corrections at
one-loop, the propagator for a particle of four-momentum $p^\mu \equiv
(E,\vec{p})$
and bare rest mass $m_0$ takes the form
\be
G = \left(\slp - m_0 -\Sigma\right)^{-1} \simeq  {\left(\slp - \Sigma_1
\slp + m_0 + \Sigma_0 \right)\over \left( p^2 - 2 \Sigma_1 p^2 - m_0^2 - 2
m_0
\Sigma_0 \right)} ~~~,
\label{2}
\ee
where $\Sigma_i=\Sigma_i(p^2)$ with $i=0,1$ are two scalar functions. Note
that in the denominator of \eqn{2} we have neglected second order terms in
$\Sigma_i$. Using standard manipulations one obtains the following
expressions for the mass shift $\delta m$ and the wave function
renormalization factor $Z_2$, respectively,
\begin{eqnarray}
\delta m & = &  \left. \Sigma \right|_{\slp=m_0} =m_0
\left. \Sigma_1 \right|_{p^2=m_0^2} +
\left. \Sigma_0 \right|_{p^2=m_0^2}
~~~,\label{covbis} \\ Z_2 & = & 1+ \left. \frac{d \Sigma}{d \slp}
\right|_{\slp=m_0} = 1 +
\left. \Sigma_1 \right|_{p^2=m_0^2}
+ 2 m_0^2 \left. \frac{d \Sigma_1}{d p^2}\right|_{p^2=m_0^2} + 2 m_0
\left. \frac{d \Sigma_0}{d p^2}\right|_{p^2=m_0^2}~~~.
\label{covariant}
\end{eqnarray}
These results can be equally obtained by expanding the propagator around
the free positive energy pole $E=\omega_p\equiv \sqrt{|\vec{p}|^2 + m_0^2}$
(for the negative energy pole one proceeds in the same way), and looking
for the perturbed value $\omega_p^R$. The residue at this pole will provide
the wave function renormalization factor and give again (\ref{covariant}).
Expanding $\Sigma_i$ up to first order in $E-\omega_p$ we get in
fact\footnote{Actually the $\Sigma_i$ should be expanded around the still
unknown value $\omega_p^R$, but the difference is of higher order in
perturbation theory.}
\be
\left( p^2 - 2 \Sigma_1 p^2 - m_0^2 - 2 m_0 \Sigma_0 \right)^{-1}
\; \simeq \;
\frac{1 + 2 \left(\wsig_1 + \wsigp_1 (E- \omega_p^R) +
\frac{m_0^2}{E+ \omega_p^R}  \wsigp_1 +  \frac{m_0}
{E+\omega_p^R}  \wsigp_0 \right) }  {(E+\omega_p^R) (E-\omega_p^R) }~~~,
\label{3}
\ee
where $\wsig_i$ and $\wsigp_i$ denote the $\Sigma_i$ functions and their
derivatives with respect to $E$ evaluated at $E=\omega_p$, which is to say,
in this covariant case, at $p^2=m_0^2$. Furthermore, $\omega_p^R$ stands
for the shifted energy pole, perturbatively close to $\omega_p$
\be
\omega_p^R \simeq \omega_p + {m_0^2 \over \omega_p} \wsig_1 +
{m_0 \over \omega_p} \wsig_0~~~,
\ee
which is equivalent to  (\ref{covbis}).

By substituting  \eqn{3} in \eqn{2} one gets for $E \simeq \omega_p^R$
\bea
G & \simeq & \frac{\left( \slp + m_0 \right)
\left( 1+ 2 \wsig_1 + \frac{ m_0^2}{\omega_p} \wsigp_1 +
 \frac{m_0}{\omega_p} \wsigp_0 \right) - \wsig_1 \slp + \wsig_0}
{2 \omega_p^R (E -
\omega_p^R) } ~~+ \mbox{finite terms} \nonumber \\ & = &
\frac{1 + \wsig_1 +  \frac{m_0^2}{\omega_p} \wsigp_1 +
 \frac{m_0}{\omega_p}  \wsigp_0}
{ E - \omega_p^R }~~
\frac{\slp + (m_0 + \delta m) }{2 \omega_p^R} +
\mbox{finite terms}~~~,
\label{4}
\eea
where now $\slp$ is understood to be $\slp = \omega_p^R \gamma_0 -
\vec{p} {\cdot} \vec{\gamma}$ and $\delta m = m_0 \wsig_1 + \wsig_0$,
which is again (\ref{covbis}). From this expression we immediately read off
the residue at the shifted positive pole $\omega_p^R$ which represents the
renormalized projector operator on positive energy states
\be
\Lambda^{+}_R = \left(1 + \wsig_1 + { m_0^2 \over \omega_p} \wsigp_1 +
{ m_0 \over \omega_p} \wsigp_0\right)
\frac{\left(\slp + m_0 + \delta m \right)}{2 \omega_p^R}
\equiv Z_2 \frac{\left(\slp + m_0 + \delta m \right)}{2 \omega_p^R}
~~~.
\label{5}
\ee
Thus, as well known, in the covariant zero temperature case, $\Lambda^+_R$
preserves the same spinor structure of the free field projector and, as
expected, the value of $Z_2$ obtained within this approach coincides with
the expression
\eqn{covariant}.

The renormalized projector operator for negative energy states
(antiparticles), $\Lambda_R^-$, is obtained from \eqn{5} by simply
replacing $(\omega_p,~ \omega_p^R) \rightarrow (-\omega_p,~ -\omega_p^R)$
and $\vec{p}
\rightarrow -\vec{p}$ (more simply $p^\mu \rightarrow -p^\mu$).
This is true since the Lorentz invariance of $\Sigma_i$ guarantees a
quadratic dependence of these quantities on $E$. Thus, $\wsig_i$ and
$\wsigp_i$ of
\eqn{5}, evaluated at the positive energy pole, are simply connected with
the same quantities evaluated at the negative energy pole, which occurs in
the expression of $\Lambda_R^-$. As we will see later, the situation can be
quite different in presence of a medium.

We now consider the finite temperature case. As before, we first expand the
propagator around the shifted positive (negative) energy pole. The residue
of $G$ at the pole will automatically give the perturbed positive
(negative) energy projector operator. In the medium rest frame, the
radiative correction $\Sigma$ of Eq. \eqn{2} takes the general form
\be
\Sigma = A(E,|\vec{p}|)~ E \gamma^0 - B(E,|\vec{p}|)~
 \vec{p} {\cdot} \vec{\gamma} - C(E,|\vec{p}|)~~~~,
\label{6}
\ee
where $A$, $B$ and $C$ are scalar functions. Note that the covariant limit
corresponds to take $A=B$ and all dependence of $A$, $B$ and $C$ on
momentum and energy via the scalar quantity $p^2$. Using the above
expression for $\Sigma$ we get for the propagator $G$
\be
G \simeq  {\left(\slp + m_0 - A E \gamma^0 + B \vec{p} {\cdot} \vec{\gamma} - C
\right) \over
\left( p^2 - 2 A E^2 - m_0^2 + 2 B |\vec{p}|^2 + 2 m_0 C \right)}~~~.
\label{7}
\ee
As for Eq. \eqn{3}, also in this case we expand $A$, $B$ and $C$ near
$E=\omega_p$, and by using the same notation we have
\bea
&& \left( p^2 - 2 A E^2 - m_0^2 + 2 B |\vec{p}|^2 + 2 m_0 C \right)^{-1}
\simeq  \nonumber \\
&& \frac{1 + 2
\left( \wa + \frac{E-\omega_p^R}{E+ \omega_p^R} \wa +
\frac{\omega_p^2}{E+\omega_p^R} \wap - \wbp {|\vec{p}|2 \over
E +\omega_p^R}- {m_0 \over E + \omega_p} \wcp\right)}{(E +
\omega_p^R) (E-\omega_p^R)} ~~~,
\label{8}
\eea
with
\be
\omega_p^R = \omega_p + \omega_p \wa - {|\vec{p}|^2 \over
\omega_p} \wb - {m_0 \over \omega_p} \wc~~~,
\label{10}
\ee
representing the dispersion relation for the finite temperature physical
state and the new positive energy pole. This result can be also written as
\be
\omega_p^R = \sqrt{|\vec{p}|^2 + (m_0 + \delta m)^2 } \simeq
\omega_p + \frac{m_0}{\omega_p} \delta m~~~,
\label{10bis}
\ee
where, from (\ref{10}), the mass shift $\delta m$ is given by
\be
\delta m = {\omega_p^2 \over m_0}\wa - {|\vec{p}|^2 \over
m_0} \wb - \wc~~~.
\label{11}
\ee
Thus for $E \simeq \omega_p^R$ the propagator reads
\be
G \simeq  \frac{1 + 2 \wa + \omega_p \wap - \wbp {|\vec{p}|^2 \over
\omega_p}- {m_0 \over \omega_p} \wcp}{2 \omega_p^R\left(E - \omega_p^R
\right)}
\left(\slp + m_0 - \wa E \gamma^0 +\wb \vec{p} {\cdot} \vec{\gamma}
- \wc \right)   + \mbox{finite terms} ~~~,
\label{9}
\ee
with $\slp = \omega_p^R \gamma_0 - \vec{p} {\cdot} \vec{\gamma}$. Using the
expression of $\delta m$ and rearranging the terms we get for the residue
at $\omega_p^R$, i.e. for the renormalized projector on positive energy
states
\be
\Lambda^+_R = \left( 1 + \wa + \omega_p \wap - {|\vec{p}|^2 \over \omega_p}
\wbp - {m_0 \over \omega_p} \wcp\right) {\left(\slp + m_R
\right) \over 2 \omega_p^R} + {\left( \wb - \wa\right) \over 2 \omega_p}
\left[ \vec{p} {\cdot} \vec{\gamma} + {|\vec{p}|^2 \over m_0}\right]~~~,
\label{12}
\ee
with $m_R = m_0 + \delta m$.

From this expression we read the two effects produced by the interactions
with the medium. First of all the zero temperature mass is shifted to the
new effective value $m_R$, which is in general momentum dependent. The
second one is a genuine contribution due to the wave function
renormalization, which consists not only of a non trivial multiplicative
factor but also introduces an additional term proportional to $\wb-\wa$ and
with a different spinorial structure. This is the peculiar signature of
breaking of Poincar\`{e} invariance which introduces a spinor dependence in
wave function renormalization. Notice that both terms are momentum (and
frame) dependent. It is easily seen that at zero temperature, i.e. in the
covariant limit, $A=B$, the second term vanishes and the first one
reproduces the usual multiplicative renormalization factor.

In order to obtain the renormalized projector on negative energy states one
should perform an analogous computation but expanding now all the involved
quantities around the negative energy pole $-\omega_p$. In the general
case, $A$, $B$ and $C$ do not have definite properties of parity with
respect to $E$ and so $\Lambda_R^+$ and $\Lambda_R^-$ can have quite
different expressions which are not simply connected one each other. This
is strongly related with the property of charge conjugation invariance of
medium and interaction. For example, for $e^-$ and $e^+$ in thermal
equilibrium with photons one only expects their chemical potentials to be
opposite. However, if the two distributions are equally populated, and thus
chemical potentials vanish, the electron-positron plasma is charge
conjugation invariant and hence $A$, $B$ and $C$ are even functions of $E$.
This is for example the situation occurring at the time of BBN. Hereafter
we will make this assumption, though more intriguing situations could be
considered as well. Under the assumption of charge conjugation invariance,
$\Lambda^-_R$ results to be
\be
\Lambda^-_R = \left( 1 + \wa + \omega_p \wap - {|\vec{p}|^2 \over \omega_p}
\wbp - {m_0 \over \omega_p} \wcp\right) {\left(\slp - m_R
\right) \over 2 \omega_p^R} + {\left( \wb - \wa\right) \over 2 \omega_p}
\left[ \vec{p} {\cdot} \vec{\gamma} - {|\vec{p}|^2 \over m_0}\right]~~~.
\label{12-}
\ee
The results of \eqn{12} and \eqn{12-} are, in a sense, half way between the
approaches of Ref.s \cite{Donoghue} and \cite{Sawyer}. The simple
multiplicative factor, in fact, is exactly the one obtained in
\cite{Donoghue}, but in that approach the additional term is absent.
Denoting with $\Lambda_{DH}^{\pm}$ the projector obtained in \cite{Donoghue} we
get
\be
\Lambda^{\pm}_{DH} = \Lambda^{\pm}_R - {\left( \wb - \wa\right) \over 2 \omega_p}
\left[ \vec{p} {\cdot} \vec{\gamma} {\pm} {|\vec{p}|^2 \over m_0}\right]~~~.
\label{13}
\ee
Actually the approach followed there is substantially different than ours.
They start introducing finite temperature spinors $\tilde{\psi}$, chosen as
the solution of the nonlinear Dirac equation
\be
(\slp - m_0 - \Sigma) \tilde{\psi}=0~~~,
\label{dirac}
\ee
whose corresponding creation and annihilation operators are assumed to
satisfy ordinary, zero temperature, anticommutation relations. Expanding
the propagator in terms of these spinors they obtain a wave function
renormalization factor which is only multiplicative. We think that the
assumption made on the {\it canonical} spinor basis to be used is
responsible for this feature and represents the essential difference with
our approach. As mentioned, we do not make any hypothesis on the
renormalized field, if any, to be used, but simply recover the particle
content and the corresponding projector operators from the poles of $G$ and
their residues, respectively.

The fact that the renormalized spinors are related to the free Dirac ones
via a momentum dependent transformation in spinor space has been first
stressed in \cite{Sawyer}. In this analysis the projector operators are
deduced, as in our approach, looking at the residue of the propagator, and
the following result is obtained, in term of our $\Lambda_R^{\pm}$
\be
\Lambda^{\pm}_S = \Lambda^{\pm}_R - {\left( \wb - \wa + {m_0 \over \omega_p}
\wcp - {m_0^2 \over \omega_p} \wbp \right) \over 2 \omega_p}
\left[ \vec{p} {\cdot} \vec{\gamma} {\pm} {|\vec{p}|^2 \over m_0}\right]~~~.
\label{14}
\ee
It is still unclear to us the reason for the difference between
$\Lambda_S^{\pm}$ and our result, since the starting point, in both cases, is
the same. It should be stressed, however, that $\Lambda_S^{\pm}$ does not
reproduce the expected behaviour in the zero temperature, covariant limit.
In fact, since we have already shown that $\Lambda_R^{\pm}$ gives back in this
limit the correct result, $\Lambda_S^{\pm}$ will still contain an explicitly
noncovariant term, only vanishing if $\wcp = m_0 \wbp$, which is not
guaranteed by any general principle.

\section{Applications to simple processes}

In order to show the physical differences of the three different approaches
to thermal wave function renormalization, simply summarized by the use of
projectors $\Lambda^{\pm}_R$, $\Lambda^{\pm}_{DH}$ and $\Lambda^{\pm}_S$, let us
discuss a relevant example like the leptonic Higgs decay $H \rightarrow l^+
l^-$. A similar analysis has been performed in \cite{Chapman}. At
tree-level the matrix element for this process reads
\be
{\cal A}(H \rightarrow l^+ l^-) = - {i g m_l \over 2 M_W} \overline{u}(l^-)
v(l^+)~~~,
\label{15}
\ee
where $m_l$ stands for the zero temperature lepton mass, and $u(l^-)$,
$v(l^+)$ denote the $T=0$ Dirac spinors for lepton and antilepton,
respectively. In a thermal bath of equally populated lepton--antilepton
pairs\footnote{We assume the Higgs mass much larger than the lepton
temperature so that the decaying particle can be considered at $T=0$.}, the
above decay width acquires an additional contribution, due to the lepton
wave function renormalization, which at one loop results to be
\bea
\left| \overline{{\cal M}} \right|^2_T &=&
\sum_{\lambda^+,\lambda^-}\left| {\cal A}(H \rightarrow l^+(\lambda^+)
l^-(\lambda^-))\right|^2  = {g^2 m_l^2 \over 4 M_W^2} \mbox{Tr}
\left[\Lambda^+(l^-) \Lambda^-(l^+) \right]
\nonumber\\
&=&{g^2 m_l^2 \over 4 M_W^2} \mbox{Tr}
\left[\left(\Lambda^+_0(l^-) + \delta\Lambda^+(l^-)\right)
\left(\Lambda^-_0(l^+)  + \delta\Lambda^-(l^+)\right)\right]
\nonumber\\
&\simeq& {g^2 m_l^2 \over 4 M_W^2} \left\{\mbox{Tr}
\left[\Lambda^+_0(l^-)\Lambda^-_0(l^+)\right] +
\mbox{Tr} \left[\delta\Lambda^+(l^-)\Lambda^-_0(l^+) +
\Lambda^+_0(l^-)\delta\Lambda^-(l^+)\right] \right\}
\nonumber\\
&=&
\left| \overline{{\cal M}} \right|^2_0
+ \delta\left| \overline{{\cal M}} \right|^2 ~~~.
\label{16}
\eea
In Eq. \eqn{16} $\lambda^{\pm}$ stand for the lepton polarizations and with $
\Lambda_0^{\pm}(l^\mp)$ and $\delta\Lambda^{\pm}(l^\mp)$ we denote the zero
temperature lepton projector and its first order thermal correction,
respectively. The second term in the r.h.s. of \eqn{16}, $\delta\left|
\overline{{\cal M}} \right|^2$, then represents the thermal contribution to
the decay width due to thermal wave function renormalization.  The
expression for $\delta\Lambda^{\pm}(l^\mp)$ can be extracted from $\Lambda^{\pm}$
of Eq.s
\eqn{12},\eqn{12-}, $\Lambda^{\pm}_{DH}$ or $\Lambda_S^{\pm}$. Once substituted in
$\delta\left|
\overline{{\cal M}} \right|^2$ we get in the c.m. reference frame
\bea
\delta\left| \overline{{\cal M}} \right|^2_R & = &
{g^2 m_l^2 \over \ M_W^2}\left( 1 - {4 m_l^2 \over M_H^2}\right)
\left[ 2 \wa - \wb  + \frac{M_H}{2} \wap - {M_H \over 2}
\left(1 - \frac{4 m_l^2}{M_H^2} \right)
\wbp- {2 m_l \over M_H} \wcp\right]~~~,\label{17}\\
\delta\left| \overline{{\cal M}} \right|^2_{DH}& = &
{g^2 m_l^2 \over \ M_W^2}\left( 1 - {4 m_l^2 \over M_H^2}\right)
\left[ \wa   + \frac{M_H}{2} \wap - {M_H \over 2}
\left(1 - \frac{4 m_l^2}{M_H^2} \right)
\wbp- {2 m_l \over M_H} \wcp\right]~~~,\label{18}\\
\delta\left| \overline{{\cal M}} \right|^2_S& = &
{g^2 m_l^2 \over \ M_W^2}\left( 1 - {4 m_l^2 \over M_H^2} \right)
\left[ \wa + {M_H \over 2} \left( \wap -\wbp \right) \right]
~~~.\label{19}
\eea
We have not included the effect due to the renormalization of the lepton
mass, in order to disentangle the two different effects. To consider its
contribution it is sufficient to substitute the value of $m_l$ and
$\omega_p$ with $m_l^R = m_l + \delta m$  and  $\omega_p^R = \omega_p +
\delta \omega$ reported in (\ref{11}) and (\ref{10bis}) respectively,
in the tree level squared amplitude
\be
\left| \overline{{\cal M}} \right|^2_0 \Rightarrow
\frac{g^2 m_l^2 }{2 M_W^2}
\left(1 - \frac{4 \left( m_l^R \right)^2}{M_H^2} \right)~~~.
\label{20}
\ee
A similar shift can also be performed in any of the contributions
(\ref{17})-(\ref{19}), since the difference we introduce in this way on the
squared amplitude is of higher order in perturbation expansion.

All approaches give the expected behaviour for the squared
amplitude\footnote{In this respect we note that the result quoted in
\cite{Chapman} as equation $(11)$ is incorrect.}, which must vanish if
$M_H=2m_l$. Nevertheless, in the expressions
\eqn{17}-\eqn{19} the thermal corrections take different forms. It is worth
observing that both our result and the one obtained using $\Lambda_{DH}^{\pm}$
coincide in the covariant limit and give the expected $Z_2$ factor reported
in \eqn{covariant}. This is not the case if $\Lambda^{\pm}_S$ is instead used.

Another example where to compute the additional contribution due to the
finite temperature renormalization wave function for electrons is the
$\beta$-decay $n \rightarrow p + e^- + \overline{\nu}_e$ and the crossed
related processes. Actually, the need for a consistent way of computing
thermal corrections to these processes, relevant for primordial
nucleosynthesis, represents one of the main motivation for our present
analysis. According to the notation adopted in \cite{Cambier} a simple
computation shows that, in the nonrelativistic limit for both neutron and
proton, the use of the renormalized projector $\Lambda_R^+$ for the
outgoing electron gives the additional contribution to the rate per neutron
\bea
\Delta \omega_S & = & \frac{ G_F^2 (g_V^2 + 3 g_A^2 ) }{2 \pi^3}
\int_0^\infty d|\vec{p}|  ~|\vec{p}|^2 Q^2 ~\theta(Q) \left(
\wa(|\vec{p}|) + \omega_p \wap(|\vec{p}|) - {|\vec{p}|^2 \over \omega_p}
\wbp(|\vec{p}|) - {m_0 \over \omega_p} \wcp(|\vec{p}|)\right) \nonumber \\
 & {\times} &  (1-F_\nu(Q))(1-F_e(\omega_p)) ~~~,
\label{betad}
\eea
where $\vec{p}$ is electron momentum, $\omega_p$ its energy,
$Q=M_n-M_p-\omega_p$ the neutrino energy and $F_{\nu,e}$ are the neutrino
and electron Fermi-Dirac distributions, respectively. The corresponding
corrections due to electron (positron) wave--function renormalization for
the other crossed processes are easily obtained replacing the thermal
factors and the expression of $Q$ as reported in Table 1 of ref.
\cite{Cambier}. Despite of the different forms of $\Lambda_R^+$,
$\Lambda_S^+$ and $\Lambda_{DH}^+$, it is interesting to note that unlike
the scalar boson decay previously considered, all give the same result
\eqn{betad} in the nonrelativistic limit. The reason is that the extra
pieces which make different the three  expressions (see Eq.s \eqn{13} and
\eqn{14}) give a correction which vanishes after the angular integration of
the differential rate is performed. This fact has been already noticed in
\cite{Chapman}.

\section{Conclusions}

In this paper we have addressed the problem of finite temperature
renormalization effects on spinor wave function. This issue has been
previously considered by many authors, with different results. The main
question seems to be whether this renormalization effect is simply
multiplicative, as in ordinary quantum field theory, or rather involves a
non trivial spinorial dependence as well. Our starting point has been to
look for the residue at the energy poles of the propagator, corrected for
thermal interactions with the surrounding medium. These poles give the new
energy--momentum dispersion relation and illustrate the particle content of
the theory. This point of view bypass the difficulty and ambiguity related
with the choice of the renormalized fields to be used at finite
temperature, which are due to the non--local and frame dependent character
of the medium effects. The projector operators on positive and negative
states we have obtained show that, as pointed out in \cite{Sawyer}, the
breaking of Poincar\'{e} invariance leads to the appearance of extra
contributions with a different spinorial structure. They reduce to ordinary
renormalized projectors in the covariant limit.

The physical interest for this problem is mainly connected with the study
of thermal corrections to nuclear interactions in the early universe,
during the primordial nucleosynthesis epoch. Notably, our results for the
case of $\beta$-decay and related processes completely agree, in the non
relativistic regime, with the ones which can be obtained using different
approaches or results for wave function renormalization, since the extra
contributions mentioned above play no role in this limit. This is not the
case for the decay of a scalar boson in fermion--antifermion pair where we
have found using our approach, a different result for the corresponding
change in the decay rate.

\end{document}